\documentclass[aps,superscriptaddress,showpacs,showkeys,nofootinbib,notitlepage,prd,preprint]{revtex4-1}
\usepackage{graphicx}
\usepackage{longtable}

\newcommand{\ie}{i.e.,\,}
\newcommand{\be}{\begin{equation}}
\newcommand{\ee}{\end{equation}}
\newcommand{\bea}{\begin{eqnarray}}
\newcommand{\eea}{\end{eqnarray}}
\newcommand{\etal}{et al.}

\newcommand{\tp}{{\tilde{\phi}_1}}

\newcommand{\sit}{\sin\tilde{\phi}_1}

\newcommand{\sitsq}{\sin^2\tilde{\phi}_1}

\newcommand{\sitcu}{\sin^3\tilde{\phi}_1}

\newcommand{\sitqu}{\sin^4\tilde{\phi}_1}

\newcommand{\cto}{\cos\tilde{\phi}_1}

\newcommand{\Lr}{\Lambda r_0^2}

\newcommand{\pht}{\tilde{\phi}_1}
\begin{document}

\bibliographystyle{apsrev}
\title{Time Delay in Swiss Cheese Gravitational Lensing }

\author{B. Chen}
\email{Bin.Chen-1@ou.edu}
\affiliation{Homer L.~Dodge Department~of  Physics and Astronomy, University of
Oklahoma, 440 West Brooks, Room~100,  Norman, OK 73019, USA}

\author{R. Kantowski}
\email{kantowski@nhn.ou.edu}
\affiliation{Homer L.~Dodge Department~of  Physics and Astronomy, University of
Oklahoma, 440 West Brooks, Room~100,  Norman, OK 73019, USA}
\author{X. Dai}
\email{xdai@ou.edu}
\affiliation{Homer L.~Dodge Department~of  Physics and Astronomy, University of
Oklahoma, 440 West Brooks, Room~100,  Norman, OK 73019, USA}
\date{\today}

\begin{abstract}
We compute  time delays  for  gravitational lensing in a flat $\Lambda$CDM Swiss cheese universe. We assume a primary and secondary pair of light rays are deflected by a single point mass condensation described by a Kottler metric (Schwarzschild with  $\Lambda$) embedded in an otherwise homogeneous cosmology. We find that the cosmological constant's effect on the difference in arrival times is  non-linear and at most around $0.002\%$ for a large cluster lens; however, we find differences from time delays predicted by conventional linear lensing theory that can reach $\sim 4\%$  for these large lenses. The differences in predicted delay times are due to the failure of conventional lensing to incorporate the lensing mass into the mean mass density of the universe.

\end{abstract}

\pacs{98.62.Sb, 95.30.Sf, 98.80.-k}

\keywords{Gravitational Lensing; General Relativity; Cosmology}

\maketitle

While investigating a recently proposed effect of  the cosmological constant $\Lambda$ on gravitational lensing \cite{Rindler07}, the authors discovered that  angular deflections caused by the embedding of point masses  in homogeneous cosmologies can differ from the Einstein value by as much as a few percent \cite{Kantowski2010}. The difference in the deflection angles is primarily caused by making the deflector's mass part of the universe's mean mass density rather than as an addition to it, as is assumed in conventional linear lensing theory. In this letter we use the same model to compute analytic expressions for time delays between images produced by these lenses. To incorporate the lenses into the mean homogeneous mass density and to enable us to evaluate time delays beyond the linear term we use a simple Swiss cheese cosmology \cite{Einstein45, Schucking54, Kantowski69a}.  Estimations of these time delays using a similar lensing model have recently appeared in \cite{Boudjemaa10, Schucker10} and earlier estimates, some using less precise models, appeared in \cite{Ishak08, Schucker08, Sereno08, Schucker09, Sereno09}. These attempts primarily focus on determining the cosmological constant's effect on lensing, including time delays,  whereas this analytic work emphasizes the significantly larger effect caused by embedding the lens into the cosmology rather than the conventional approach of simply superimposing  the lensing mass on top of the homogeneous mean density.

Swiss cheese models are exact general relativistic (GR) solutions which replace comoving homogeneous spheres in Friedman-Lema\^itre-Robertson-Walker (FLRW) universes by spherical condensations (lenses) \cite{Kantowski69b}. Because a Swiss cheese lens is part of the mean density there is a range beyond which it ceases to deflect passing light rays. That range is just the comoving radial boundary $\chi_b$ of the homogeneous sphere that is replaced by the condensation. Beyond that distance the gravity caused by a condensation and a homogeneous sphere are the same.

\begin{figure*}
\includegraphics[angle=90,width=0.7\textwidth,height=0.28\textheight]{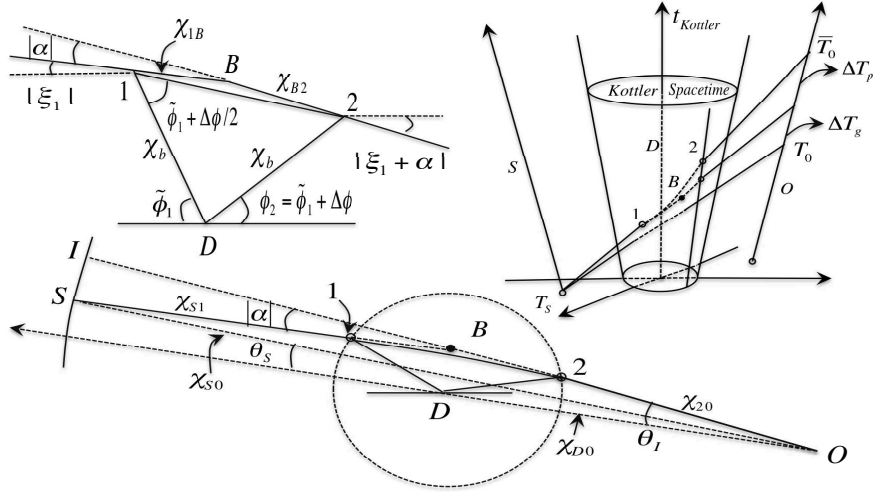}
\caption{Swiss cheese gravitational lensing. Bottom: the  spatial comoving paths of three photons. The deflected photon leaves a source $S$, enters a Kottler hole at point 1, exits at 2 at a deflected angle $\alpha$, and then proceeds to the observer at 0. A second photon travels straight from the source to the observer as if the Kottler condensation were absent. An imagined third photon also travels as if the condensation were absent but is reflected at point B by angle $\alpha$ before arriving at 0. The point $B$ (the reflection point) is the intersection of the forward and backward extensions of respective FLRW rays ${S1}$ and ${20}$ drawn as if the  Kottler hole were absent. $\chi_{S1}$ and  $\chi_{20}$ are comoving  distances respectively from the source $S$ to entrance point 1 and from exit point 2 to the observer.  $\chi_{S0}$ and  $\chi_{D0}$ are  comoving distances from the source and deflector respectively to the observer. The comoving distances from source to reflection point B and from B to the observer are respectively $\chi_{SB}\equiv \chi_{S1}+\chi_{1B}$ and $\chi_{B0}\equiv\chi_{B2}+\chi_{20}.$  The angular positions (as seen by the observer) of the image and source relative to the optical axis are respectively $\theta_I$ and $\theta_S$. Top left: a blow up of the two triangles $1B2$ and $1D2$. The angles $\tp, \xi_1, \alpha,$ and $\Delta\phi$  are described below Eq.\,(\ref{chi-1B2}) and are the same as those computed in \cite{Kantowski2010}. Top right: a space time diagram showing the difference in arrival times $\Delta T_g$ and $\Delta T_p$ of three photons originating from source S at time $T_S$ and arriving at observer $0$. The upper photon encounters a Kottler condensation at time $T_1$, exits at $T_2$, and then arrives at the observer at time $\overline{T}_0=T_0+\Delta T_g+\Delta T_p$. The lower photon arrives at $T_0$ after traveling on a straight line entirely in FLRW. The middle photon arrives at time $T_0+\Delta T_g$ after traveling on two straight segments ${SB}$ and ${B0}$ both entirely in FLRW but whose directions differ by the angle $\alpha$.}
\label{fig:outside}
\end{figure*}

In this calculation we assume the background FLRW cosmology is a flat ($\Omega_{\rm m}+\Omega_\Lambda=1$) $\Lambda$CDM model. The simplest Swiss cheese cosmology is constructed by replacing  the co-moving spheres of homogeneous dust by point masses \cite{Einstein45,Schucking54}. When $\Lambda\ne 0$ the metric in an evacuated void is the Kottler metric \cite{Kottler18,Dyer74}  (Schwarzschild with a cosmological constant present).
In \cite{Kantowski2010} we calculated in detail the deflection angle of a photon passing through a Kottler condensation. We now compute  analytic expressions for the time delay caused by encountering such a deflector. We compute the difference in arrival times $\Delta T_0$ of two light rays emitted at the same time from two sources at equal comoving distances from the observer. One ray is assumed to travel entirely in FLRW and arrive at time $T_0$ and the other ray is assumed to encounter a deflector and arrive at the observer at the later time $\overline{T}_0=T_0+\Delta T_0$ (see Fig.\,\ref{fig:outside}).

The Kottler metric \cite{Kottler18} can be written as
\be\label{Kottler}
ds^2=-\gamma(r)^{-2}c^2dt^2+
\gamma(r)^2dr^2+
r^2(d\theta^2+\sin^2\theta\, d\phi^2),
\ee
where $\gamma^{-1}(r)\equiv\sqrt{1-\beta^2(r)}$,
\be
\beta^2(r)=\frac{r_s}{r}+\frac{\Lambda r^2}{3},
\label{beta}
\ee
and $r_s= 2G{\rm m}/c^2$ is the Schwarzschild radius of the condensed mass. The flat FLRW metric for the background cosmology can be written as
\be
ds^2=-c^2dT^2+R(T)^2\left[{d\chi^2}+\chi^2(d\theta^2+\sin^2\theta d\phi^2)\right],
\label{RW}
\ee
with its time development determined by
\be
\frac{\dot{R}}{R}=H(R)\equiv H_0\sqrt{\Omega_\Lambda+\Omega_{\rm m}\left(\frac{R_0}{R}\right)^3}.
\label{Hubble}
\ee
To satisfy the junction conditions of GR the boundary  of the comoving sphere of dust removed [$\chi_b$\,=\,constant; $T,\theta,\phi$\,=\,arbitrary] is matched to an expanding sphere of radial timelike geodesics [$(t_b(T),r_b(T))$ with $\theta,\phi$ arbitrary] in the Kottler vacuole, by requiring
\be\label{boundary}
r_s=\Omega_{\rm m}\frac{H_0^2}{c^2}(R_0\chi_b)^3,\ \  r_b(T)=R(T)\chi_b.
\ee
A consequence of Eq.\,(\ref{boundary}) for flat FLRW is that the net dust removed  equals the condensed Schwarzschild mass.

\begin{figure*}
\begin{center}$
\begin{array}{cc}
\includegraphics[width=0.5\textwidth,height=0.25\textheight]{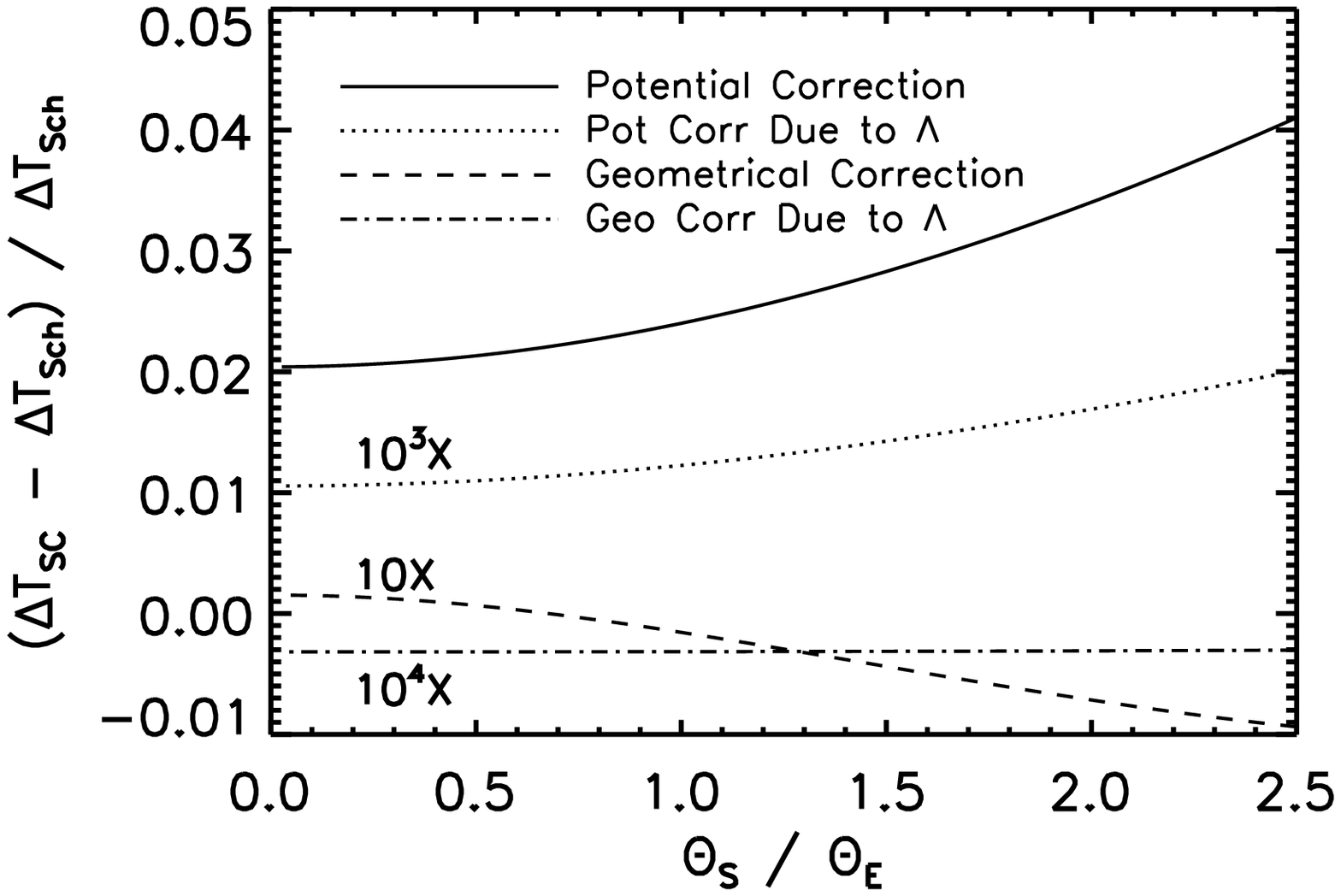}
\hspace{10pt}
\includegraphics[width=0.5\textwidth,height=0.25\textheight]{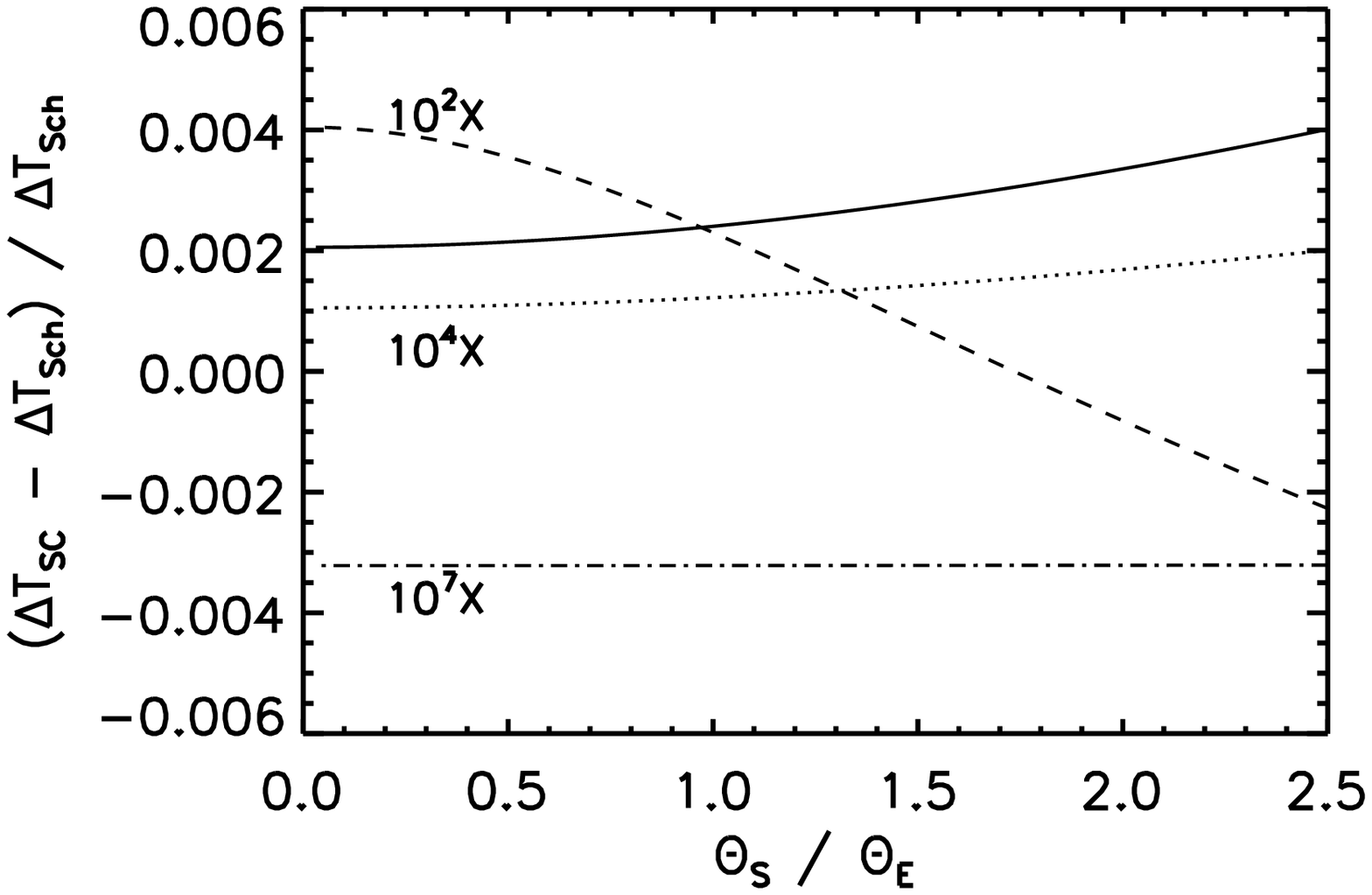}
\end{array}$
\end{center}
\caption{Fractional differences between Swiss cheese ($\Delta T_{\rm SC}$) and conventional Schwarzschild ($\Delta T_{\rm Sch}$) time delays as functions of source position $\theta_S$ in units of Einstein ring radius $\theta_E$ for a $10^{15} M_{\odot}$ (left) and a $10^{12} M_{\odot}$ (right) lens.
The solid and dashed curves are the fractional time delay differences in the potential  and the geometrical parts respectively. The dotted and dot-dashed curves are those parts of the fractional differences due to the square root term (contains $\Lambda$) in Eqs.\,(\ref{DeltaTp}) and (\ref{DeltaTg}).
}
\label{fig:M15-full}
\end{figure*}

A photon (see Fig.\,\ref{fig:outside}) traveling in a flat  $\Lambda$CDM universe starting from a source at time $T_S$ and comoving distance $\chi_{S0}$ from the observer, moves along a null geodesic of Eq.\,(\ref{RW}), and arrives at  time $T_0$ where
\be
\chi_{S0}=\int^{T_0}_{T_S}\frac{c\,dT}{R(T)}=\int^{R_0}_{R_S}\frac{cdR}{R^2H}.
\label{s0}
\ee
We want to compare this straight path travel time to the time $\overline{T}_0-T_S$ it takes a  photon to travel along  a deflected path from  the source through  a Kottler  condensation and then to the observer. If the entrance time into  and exit time from the Kottler hole are respectively $T_1$ and $T_2$, the  respective comoving distances in FLRW traveled by the deflected photon before and after encountering the Kottler hole are
\be
\chi_{S1}=\int^{R_1}_{R_S}\frac{cdR}{R^2H},\ {\rm and }\
\chi_{20}=\int^{\overline{R}_0}_{R_2}\frac{cdR}{R^2H}.
\label{s1-20}
\ee
Combining these with Eq.\,(\ref{s0}) we have an equation to solve for the difference in travel times
\be
\int^{\overline{T}_0}_{T_0}\frac{cdT}{R}=\Delta\chi\equiv\chi_{S1}+\int^{R_2}_{R_1}\frac{cdR}{R^2H}+\chi_{20}-\chi_{S0}.
\label{exact}
\ee
$\Delta\chi$ is just the difference in the comoving distance the deflected photon would have traveled if it went straight in FLRW for the actual travel time $\overline{T}_0-T_S$ and the comoving distance from the source to the observer (a travel time of $T_0-T_S$ if the deflector were not encountered). The difference in comoving distances  $\Delta\chi$ can be factored into the sum of a geometrical part and a potential part respectively  $\Delta\chi_g$ and $\Delta\chi_p$ by adding and subtracting the sum of two comoving distances $\chi_{1B}+\chi_{B2}$ (see Fig.\,\ref{fig:outside}),
\be
\Delta\chi_g\equiv \chi_{SB}+\chi_{B0}-\chi_{S0}, \ \ \Delta\chi_p \equiv \int^{R_2}_{R_1}\frac{cdR}{R^2H}-(\chi_{1B}+\chi_{B2}).
\label{Delta-chi}
\ee
The path SB0 is that of an imaginary photon that starts from the source S and travels to point 1, just as the deflected photon does, but then continues on in FLRW until it is reflected at point B by an angle $\alpha$ (same as the Kottler deflection angle) and finally arrives at the observer at $T_0+\Delta T_g$. The deflected photon, if traveling straight, would go an additional comoving distance $\Delta\chi_p$ and consequently arrives an additional time $\Delta T_p$ later.  $\Delta\chi_p$ is the difference in the comoving distance a photon would have traveled in FLRW during the Kottler crossing time $T_2-T_1$ minus the comoving distance the imaginary reflected photon did travel in crossing the un-evacuated sphere.
We identify $\Delta T_g$ and $\Delta T_p$ as the geometrical and potential parts of the time delay for purposes of comparison with standard linear lensing delays \cite{Cooke75}. A decomposition is not unique but this one is useful and results in Eqs.\,(\ref{DeltaTg}) and (\ref{DeltaTp}).

To proceed further we must relate comoving distances $\chi_{SB}+\chi_{B0}$ and $\chi_{1B}+\chi_{B2}$ to known  deflector-observer and source-observer distances $\chi_{D0}$ and $\chi_{S0}$ as well as evaluate the integral in Eq.\,(\ref{Delta-chi}).
We can avoid computing $\chi_{SB}$ and $\chi_{B0}$ to all but the lowest order $\chi_{SB}\approx \chi_{S0}-\chi_{D0}$ and $\chi_{B0}\approx \chi_{D0}$ by using the law of cosines on the triangle SB0. To the accuracy we give $\Delta T_0$ it suffices to replace the left hand side of Eq.\,(\ref{exact}) by $c\Delta T_0/R_0$.
The resulting geometrical part of the time delay is
\bea
\Delta T_g &\approx&\frac{R_0}{c}\Delta\chi_g = \frac{R_0}{c}\left(\sqrt{\chi^2_{S0}+4\chi_{SB}\chi_{B0}\sin^2(\alpha/2)}-\chi_{S0}\right)\cr
&=&\frac{R_0}{c}\frac{(\chi_{S0}-\chi_{D0})\chi_{D0}}{2\ \chi_{S0}}\alpha^2\left[1+{\cal O}(\alpha^2)+{\cal O}\left(\beta_1\frac{\chi_b}{\chi_{S0}}\right)^2\right]\cr
&\approx&\ 2 \frac{(1+z_D)}{c}\frac{D_{SD}D_{D0}}{D_{S0}}\left(\frac{r_s}{r_0}\right)^2\cos^6\tilde{\phi}_1\cr
& &\times\left(1+6\tan\pht\sqrt{\frac{\Lr}{3}+\frac{r_s}{r_0}\,\sitcu}\right),
\label{DeltaTg}
\eea
where the deflection angle $\alpha$ has been taken from Eq.\,(32) of \cite{Kantowski2010}. The source-deflector-observer distances in Eq.\,(\ref{DeltaTg}) are the usual angular diameter distances. For small impact angles $\tp$ this result reduces to the standard lensing result given in Eq.\,(12) of \cite{Cooke75}. For larger impact angles the difference is significant, even vanishing when $\tp\rightarrow\pi/2$ as the photon grazes the Kottler void.

To evaluate the potential part of the time delay in Eq.\,(\ref{Delta-chi}) we need to determine $\chi_{1B}+\chi_{B2}$. If we apply the law of sines and cosines to the triangles 1B2 and  1D2 respectively, the result is 
\bea
\chi_{1B}&+&\chi_{B2}=2\chi_b\cos(\pht+\Delta\phi/2)\cos(\Delta\phi/2-\xi_1-\alpha/2)/\cos(\alpha/2).
\label{chi-1B2}
\eea
The deflected photon's trajectory angles in this expression are shown in Fig.\,\ref{fig:outside} and are the same as those used in \cite{Kantowski2010}. They are defined as follows: In the plane of the deflected photon (spherical polar angle $\theta=\pi/2$) the incoming photon's slope is $\tan\xi_1$, the photon impacts the Kottler void at point 1 with azimuthal angle $\pi-\tp$,  the photon exits the void at point 2 with slope $\tan(\xi_1+\alpha)$ and azimuthal angle $\tp+\Delta\phi$. By definition the angles $\alpha,\, \xi_1,$ and $\Delta\phi$ are negative. Equation (\ref{chi-1B2}) can be evaluated using values for these angles given in \cite{Kantowski2010} as series that depend on the deflector's mass, the cosmological constant, the photon's minimum Kottler radius $r_0$, and the photon's entrance angle $\tp$. We evaluate the integral in Eq.\,(\ref{Delta-chi}) by expanding it as a series in $\Delta R=R_2-R_1$ to order $(\Delta R)^4$ using Eq.\,(\ref{Hubble}) where  $\Delta R$ is the change in the radius of the universe that occurs while the photon traverses the Kottler condensation. It is related to the change in the Kottler coordinate of the boundary $\Delta r=r_b(T_2)-r_b(T_1)$ that occurs during the photon's transit by Eq.\,(\ref{boundary}), \ie
\be
\Delta R=\frac{R_1}{r_1}\Delta r.
\ee
The change $\Delta r$ is related to the change in the azimuthal angle $\Delta\phi=\phi_2-\tp$ by the orbit equation (11) of \cite{Kantowski2010} and $\Delta\phi$ is given by Eq.\,(\ref{Delta-phi}) below.
In Eq.\,(13) of \cite{Kantowski2010} we gave an expression for $\Delta\phi$ but not to the accuracy needed to evaluate non-linear corrections to the time delay. Using the method described in \cite{Kantowski2010} we computed $\Delta\phi$ to the next higher order,
\bea
\Delta\phi &=&-2\beta_1 \sit +\left(\frac{r_s}{r_0}\right)\Biggl[3\cto\sitsq -\beta_1\Bigg(2+\frac{7}{3}\sitsq\nonumber\\
&&-6\sitqu+2\log\left\{\cot\frac{\tilde{\phi}_1}{2}\right\}\tan\tilde{\phi}_1\sit\Bigg)\Biggr]-
\frac{1}{9}\,\beta_1\,\Lr\sit\nonumber\\
&&+\left(\frac{r_s}{r_0}\right)^2\Biggl[2 \log \left\{\cot \frac{\tp }{2}\right\} \sin\tp  \left(4-3 \sin ^2\tp\right) \tan^2\tp\nonumber\\
 &&+\frac{1}{6} \Bigg(36 -9 \sin \tp -70 \sin ^2\tp +9 \sin ^3\tp -59 \sin ^4\tp   \nonumber\\
&&+81 \sin ^6\tp\Bigg) \tan \tp \Biggr]+\frac{r_s}{r_0}\Lambda r_0^2 \Bigg[\frac{2}{3} \log \left\{\cot\frac{\tp }{2}\right\} \sec ^2\tp\nonumber\\
&&-\frac{2}{9} \sec \tp  \left(1+14\sin ^2\tp -12 \sin ^4\tp  \right)\Bigg]+{\cal O}(\beta_1^5).
\label{Delta-phi}
\eea
In this expression $\beta_1$, see Eq.\,(\ref{beta}), is the expansion velocity of the Kottler boundary as seen by a static observer at the instant the photon enters the void.
When the above is inserted into the integral in Eq.\,(\ref{Delta-chi}) and combined with Eq.\,(\ref{chi-1B2}) the potential part of the time delay can be approximated by using $\alpha$ from Eq.\,(32) and $\xi_1$ from Eq.\,(18) of \cite{Kantowski2010}. The result is
\bea\label{DeltaTp}
\Delta T_p &=& 2\frac{(1+z_d)}{c}r_s\Biggl[ \log\left\{\cot\frac{\pht}{2}\right\}-\cos\pht\left(1+\frac{1}{3}\cos^2\pht\right)\nonumber\\
+&\cot\pht&\cos^3\pht\sqrt{\frac{\Lr}{3}+\frac{r_s}{r_0}\,\sitcu}
+{\cal O}\left(\beta_1^2+\beta_1\frac{\chi_b}{\chi_{d}}\right)\Biggr].
\eea Arriving at Eq.\,(\ref{DeltaTp}) also required converting the redshift $z_1$ of the photon when just entering the Kottler void to the slightly smaller redshift  $z_d$ of a source located at the deflector's distance. To the accuracy needed here
\be
(1+z_1)=(1+z_d)[1+\beta_1\cos\pht+{\cal O}(\beta_1^2+\beta_1\chi_b/\chi_{d})].
\ee
The result for $\Delta T_p$ in Eq.\,(\ref{DeltaTp}) can be compared with the original linear result given in Eq.\,(17) of \cite{Cooke75}. The Swiss cheese result reproduces the original result when $\tp\sim 0$ (but $\ne0 $). The comparison requires two images seen at small $\tp$'s.  The potential part of the time delay also vanishes as it should when $\tp\rightarrow \pi/2 $, \ie when the photon  misses the Kottler hole. The most significant part of the correction is to the linear term and just as with the geometrical term the cosmological constant only appears as a part of the expansion velocity of the Kottler boundaries [the square root  terms in Eqs.\,(\ref{DeltaTg}) and (\ref{DeltaTp}) come from $\beta_1$].

In Fig.\,\ref{fig:M15-full} we illustrate the use of Eqs.\,(\ref{DeltaTg}) and (\ref{DeltaTp}) by comparing conventional and Swiss cheese time delays for two point mass lens systems with $z_d=0.5,$ $z_s=1.0$ and  masses  ${\rm m}=10^{12} \>M_{\odot}$ and $10^{15} \>M_{\odot}$. The cosmological parameters used  are $\Omega_{\rm m}=0.3,$  $\Omega_\Lambda=0.7$ and $H_0=70\, {\rm km}\, {\rm s}^{-1}\, {\rm Mpc}^{-1}.$  We graph the fractional difference in time delays between Swiss cheese and  standard Schwarzschild lensing as functions of source position $\theta_S$.  For both lenses the abscissa $\theta_S$ varies from $0$ to $2.5\,\theta_E$ where $\theta_E$ is the classical Einstein ring radius. For the $10^{15}\>M_{\odot}$ case the total relative correction to the potential part of the time delay is seen to  be  larger than $2\%$ ranging up to $4\%$ at 2.5 $\theta_E$.  The effect of  $\Lambda$  is to increase the potential time delay by as much as $0.002\%.$  The geometrical part of the time delay differs from the conventional Schwarzschild result by as much as $0.1\%$ of which a reduction by a factor of $3\times 10^{-7}$ was caused by the $\Lambda$ terms in Eq.\,(\ref{DeltaTg}). For the $10^{12}\>M_{\odot}$ deflector we observe a change in the potential  part of the time delay of up to $0.4\%.$ The $\Lambda$ term inceases the potential part of the time delay by as much as $2\times 10^{-7}.$  The geometrical part of the time delay is different by as much as $4\times 10^{-5}$ whereas $\Lambda$'s effect is a reduction of no more than a factor of $3\times 10^{-10}.$

For currently observed strong lensing by clusters the net time delay corrections are only $\sim$0.2\%.
For example a pair of images for a lens like SDSS J1004+4112 at redshifts $z_d=0.68$,  $z_s=1.73$, a  lens mass of $M = 2.3\times 10^{13}$ and a source position $\theta_s=0.15\,\theta_E$, produces a Schwarzschild time delay of $\sim$7.26 yrs with a Swiss cheese increase of $\sim 4.9$ days. As indicated by  Fig.\,\ref{fig:M15-full} future lensing configurations could yield corrections 10 times larger. However, exactly how important Swiss cheese lensing is for lens modeling awaits additional theoretical work on the embedding  of more realistic distributed mass lenses.

Work on this project was partially supported by NSF grant AST-0707704 and US DOE
Grant DE-FG02-07ER41517 and B. Chen wishes to thank the University of Oklahoma Foundation for a fellowship.

\label{lastpage}

\end{document}